\documentclass[useAMS,usenatbib]{mn2e}

\usepackage{aas_macros}
\usepackage{graphicx}
\usepackage{amsmath}
\usepackage{xspace}
\usepackage{ifthen}

\newcommand\T{\rule{0pt}{4ex}}       % Top strut
\newcommand\B{\rule[-1.2ex]{0pt}{0pt}} % Bottom strut

\newcommand{\chisq}{$\chi ^2$\xspace}

\title[Secondary eclipses of WASP-19b and WASP-43b]{$K_s$ band secondary eclipses of WASP-19b and WASP-43b with the Anglo-Australian Telescope
\thanks{Based on observations obtained at the Anglo-Australian Telescope, Siding
Spring, Australia.}}

\author[G.~Zhou et al.]
{\parbox{\textwidth}
{G.~Zhou$^{1}$\thanks{E-mail: \texttt{george.zhou@anu.edu.au}},
D.D.R.~Bayliss$^{1}$,
L.~Kedziora-Chudczer$^{2,3}$,
G.~Salter$^{2,3}$,
C.G.~Tinney$^{2,3}$,
and J.~Bailey$^{2,3}$
\vspace{0.4cm}}\\
\parbox{\textwidth}{
$^{1}${Research School of Astronomy and Astrophysics, Australian National University, Canberra, ACT 2611, Australia}\\
$^{2}${School of Physics, University of New South Wales, Sydney, NSW 2052, Australia}\\
$^{3}${Australian Centre for Astrobiology, University of New South Wales, Sydney, NSW 2052, Australia}\\
}}

\begin{document}
%\linenumbers

\date{Accepted 2014 September 9.  Received 2014 September 9; in original form 2014 July 17}

\pagerange{\pageref{firstpage}--\pageref{lastpage}} \pubyear{2014}

\maketitle

\label{firstpage}

\begin{abstract}
We report new $K_s$ band secondary eclipse observations for the hot-Jupiters WASP-19b and WASP-43b. Using the IRIS2 infrared camera on the Anglo-Australian Telescope (AAT), we measured significant secondary eclipses for both planets, with depths of $0.287_{-0.020}^{+0.020}$\,\% and $0.181_{-0.027}^{+0.027}$\,\% for WASP-19b and WASP-43b respectively. We compare the observations to atmosphere models from the VSTAR line-by-line radiative transfer code, and examine the effect of C/O abundance, top layer haze, and metallicities on the observed spectra. We performed a series of signal injection and recovery exercises on the observed light curves to explore the detection thresholds of the AAT+IRIS2 facility. We find that the optimal photometric precision is achieved for targets brighter than $K_\text{mag} = 9$, for which eclipses as shallow as 0.05\,\% are detectable at $>5\,\sigma$ significance. 
\end{abstract}


\section{Introduction}
\label{sec:introduction}

Observations of secondary eclipses measure the emergent thermal flux of transiting exoplanets. These observations have provided the first glimpse into the structure and composition of hot-Jupiter atmospheres. Around 40 exoplanets have now been observed in eclipse, with the vast majority of these observations obtained from space-based observatories such as \emph{Spitzer} in the infrared \citep[e.g.][]{2005Natur.434..740D,2005ApJ...626..523C,2008ApJ...686.1341C} and \emph{Kepler} in the optical \citep[e.g.][]{2009Sci...325..709B,2011ApJS..197...11D,2012AJ....143...39C}. Ground-based observations have also been relatively successful \citep[e.g.][]{2009A&amp;A...493L..35D,2009A&amp;A...493L..31S,2009ApJ...707.1707R,2009A&amp;A...506..359G}, but are intrinsically difficult due to a combination of instrument and atmosphere-induced systematics, and remains limiting. Ground-based observations characterise eclipses at shorter wavelengths ($<2.1\,\mu\text{m}$) than the $Spitzer$ measurements, probing deeper into the atmosphere and providing a longer wavelength baseline to constrain the atmosphere models. The importance of these near infrared measurements has lead to a number of recent surveys targeting most irradiated hot-Jupiters, and making a significant contribution to the sample of planets studied in eclipse \citep[e.g.][]{2012ApJ...744..122Z,2012ApJ...748L...8Z,2013ApJ...770...70W,2014A&amp;A...564A...6C,2014A&amp;A...567A...8C,2014ApJ...788...92S}.

In this paper, we provide new observations for the eclipses of two short period, highly irradiated hot-Jupiters -- WASP-19b and WASP-43b. These are the first results from our programme at the AAT, aimed at providing a set of self consistent secondary eclipse measurements for hot-Jupiters.

WASP-19b \citep{2010ApJ...708..224H} is a $1.17\,M_\text{Jup}$, $1.39\, R_\text{Jup}$ hot-Jupiter in a 0.79 day orbit about a $K_\text{mag} =10.48$ G-dwarf. It has been well studied in eclipse, with ground-based broadband observations at the 0.67\,$\mu$m ASTEP band \citep{2013A&amp;A...553A..49A}, $i$ band \citep{2013MNRAS.436....2M}, $z$ band \citep{2012ApJS..201...36B,2013A&amp;A...552A...2L,2013ApJ...774..118Z}, $H$ band \citep{2010A&amp;A...513L...3A}, \emph{Spitzer} 3.6, 4.5, 5.8, and 8.0\,$\mu$m bands \citep{2013MNRAS.430.3422A}, narrowband detections at 1.190\,$\mu$m \citep{2013A&amp;A...552A...2L} and 2.095\,$\mu$m \citep{2010MNRAS.404L.114G}, as well as eclipse spectrophotometry over 1.25--2.35\,$\mu$m \citep{2013ApJ...771..108B}. In addition, transmission spectrophotometry of the planet was obtained from the ground by \citep{2013ApJ...771..108B}, and via the \emph{Hubble Space Telescope (HST)} by \citep{2013MNRAS.434.3252H,2013ApJ...779..128M}. The broadband emission spectrum of WASP-19b is consistent with an atmosphere lacking a thermal inversion layer \citep{2013MNRAS.430.3422A}, and marginally prefers carbon-rich atmosphere models \citep{2012ApJ...758...36M}. The detection of water in the \emph{HST} transmission spectra, however, supports a lower C/O ratio and a depletion of TiO in the atmosphere \citep{2013MNRAS.434.3252H}.

WASP-43b \citep{2011A&amp;A...535L...7H} is another well studied hot-Jupiter with mass of $1.8\,M_\text{Jup}$ and radius of $0.9\,R_\text{Jup}$, orbiting its $K_\text{mag}=9.27$ K-dwarf host star once every 0.81 days. Eclipses of the planet have been detected in $i$ band \citep{2014A&amp;A...563A..40C}, 1.19\,$\mu$m narrowband \citep{2012A&amp;A...542A...4G}, $H$ band \citep{2013ApJ...770...70W}, 2.09\,$\mu$m narrowband \citep{2012A&amp;A...542A...4G}, $K$ band \citep{2013ApJ...770...70W,2014A&amp;A...563A..40C}, and the \emph{Spitzer} 3.6 and 4.5\,$\mu$m bands \citep{2014ApJ...781..116B}. These observations can rule out the presence of a strong thermal inversion layer \citep{2013ApJ...770...70W,2014ApJ...781..116B}, but cannot place constraints on the composition of the atmosphere.

Details of the observations, data reduction, light curve extraction and analysis can be found in Section~\ref{sec:observations}. Since both targets have previous observations in similar wavelengths, Section~\ref{sec:results} compares our results to literature measurements to investigate the repeatability and robustness of secondary eclipse observations. We compare the measured emission spectra of the two planets to VSTAR atmosphere models in Section~\ref{sec:vstar-atmosph-models}. In Section~\ref{sec:iris2-detect-limits}, we perform signal injection and recovery on the residual light curve of our WASP-19 observation to understand the detection thresholds and potentials of the AAT secondary eclipse programme. Section~\ref{sec:iris2-detect-limits} also provides an estimate for the number of known planets with eclipses observable by the AAT.

\section{Observations and Analysis}
\label{sec:observations}

\subsection{Observing Strategy}
\label{sec:observing-strategy}

\begin{figure*}
  \centering
  \begin{tabular}{cc}
    \includegraphics[width=9cm]{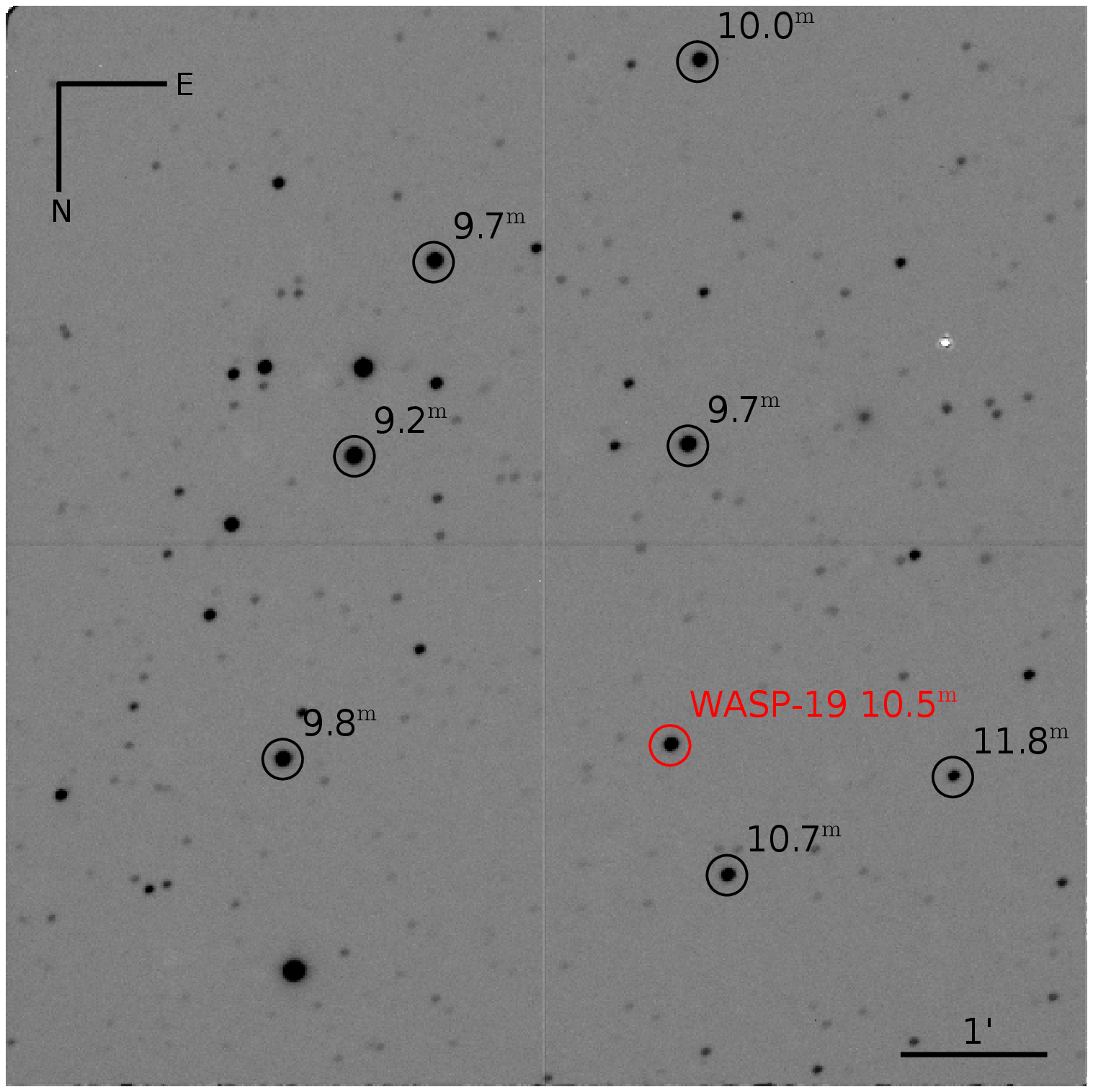} &
    \includegraphics[width=9cm]{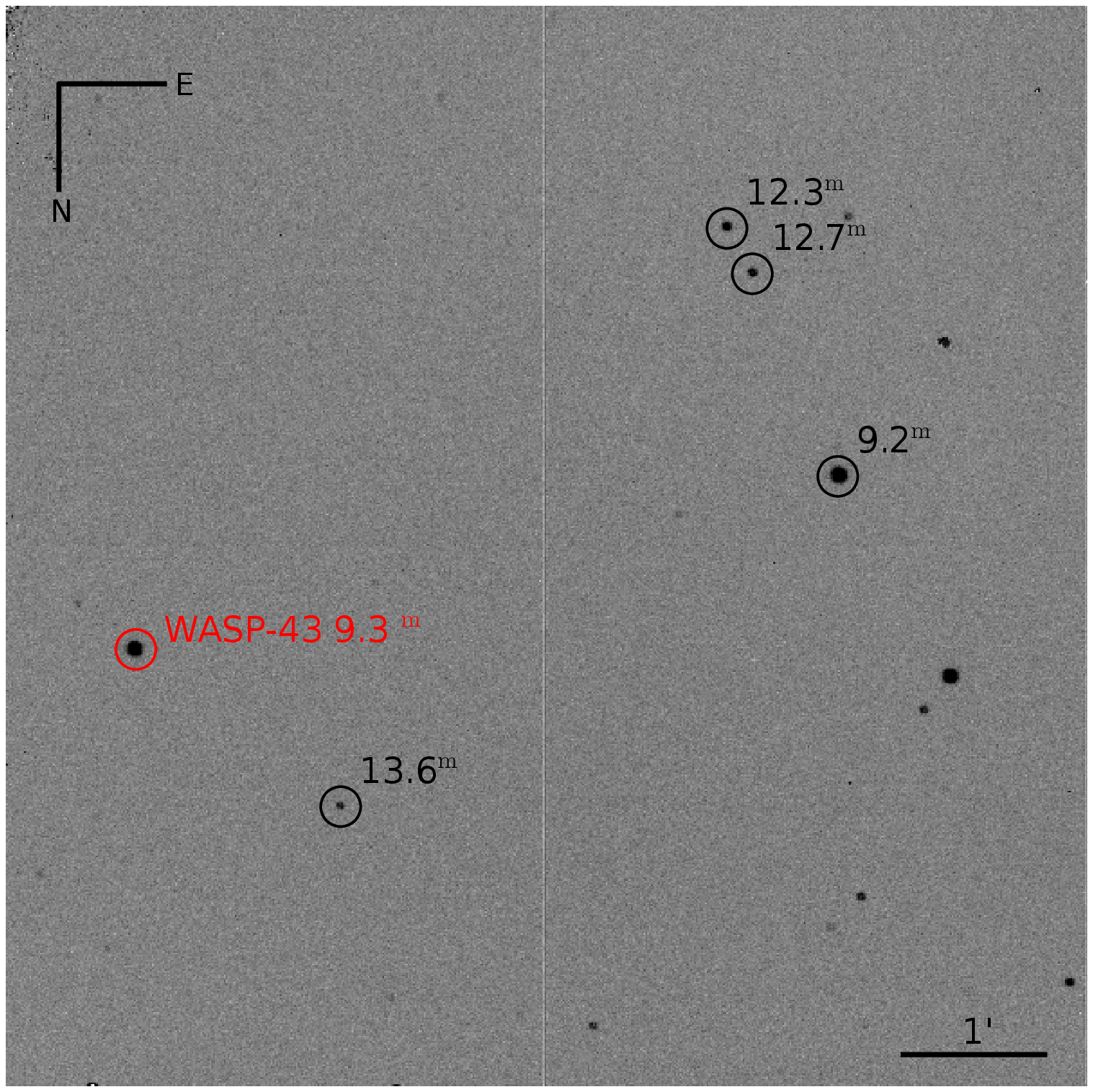}
  \end{tabular}
  \caption{Sample IRIS2 images of the WASP-19 (left) and WASP-43 (right) fields. The respective target stars are labelled in red. The reference stars used are marked in black, with their $K$ band magnitudes marked.}
  \label{fig:field_img}
\end{figure*}

\begin{figure*}
  \centering
  \includegraphics[width=15cm]{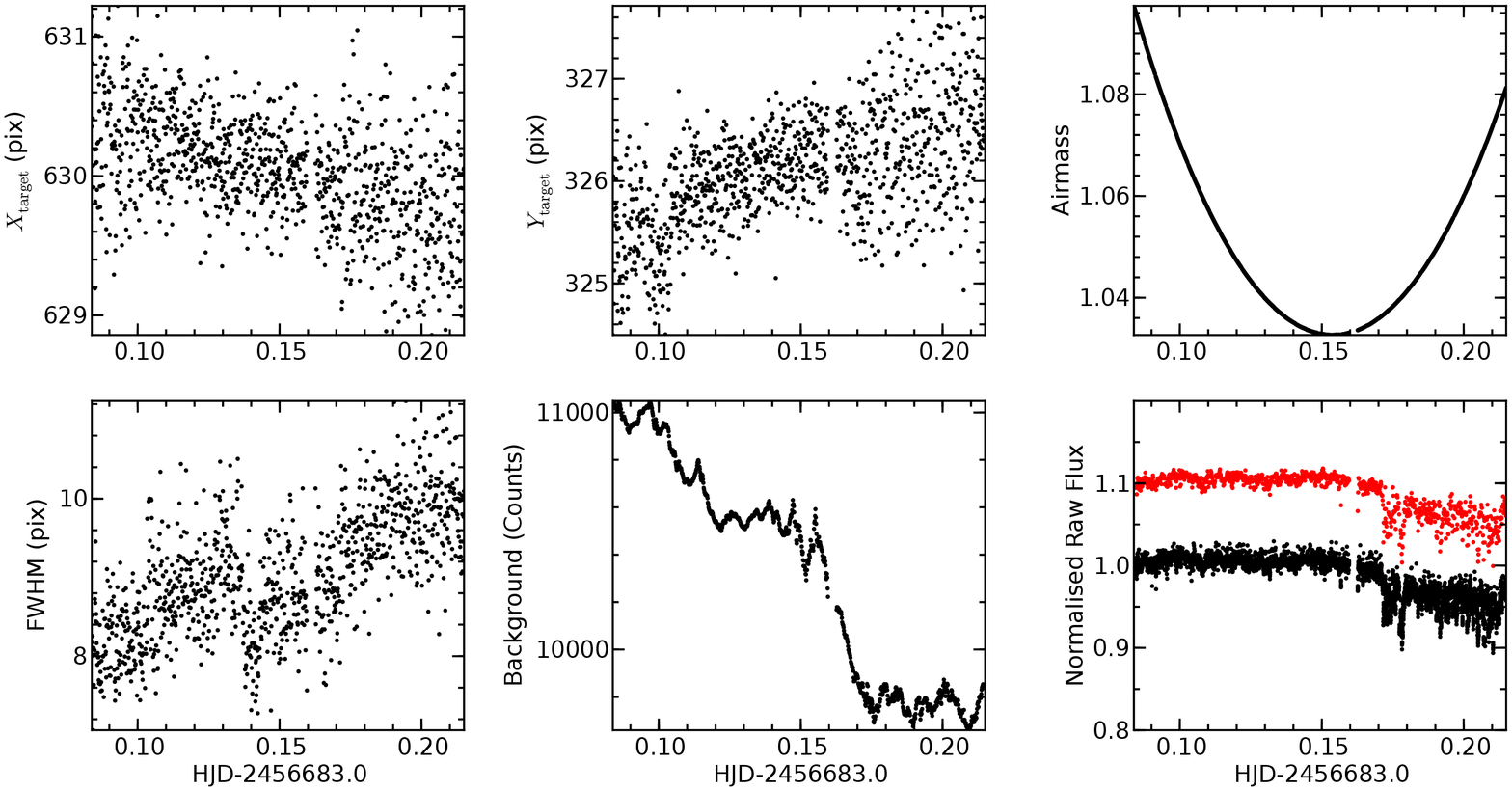}
  \caption{External parameters from the WASP-19b eclipse observation on 2014 January 25, including the target star detector $X$, $Y$ centroid positions, airmass, image PSF FWHM, sky background counts, and normalised raw fluxes for the ensemble of reference (black), and target (red, arbitrarily shifted upward by 0.1 for clarity) light curves.}
  \label{fig:W19_extern}
\end{figure*}

\begin{figure*}
  \centering
  \includegraphics[width=15cm]{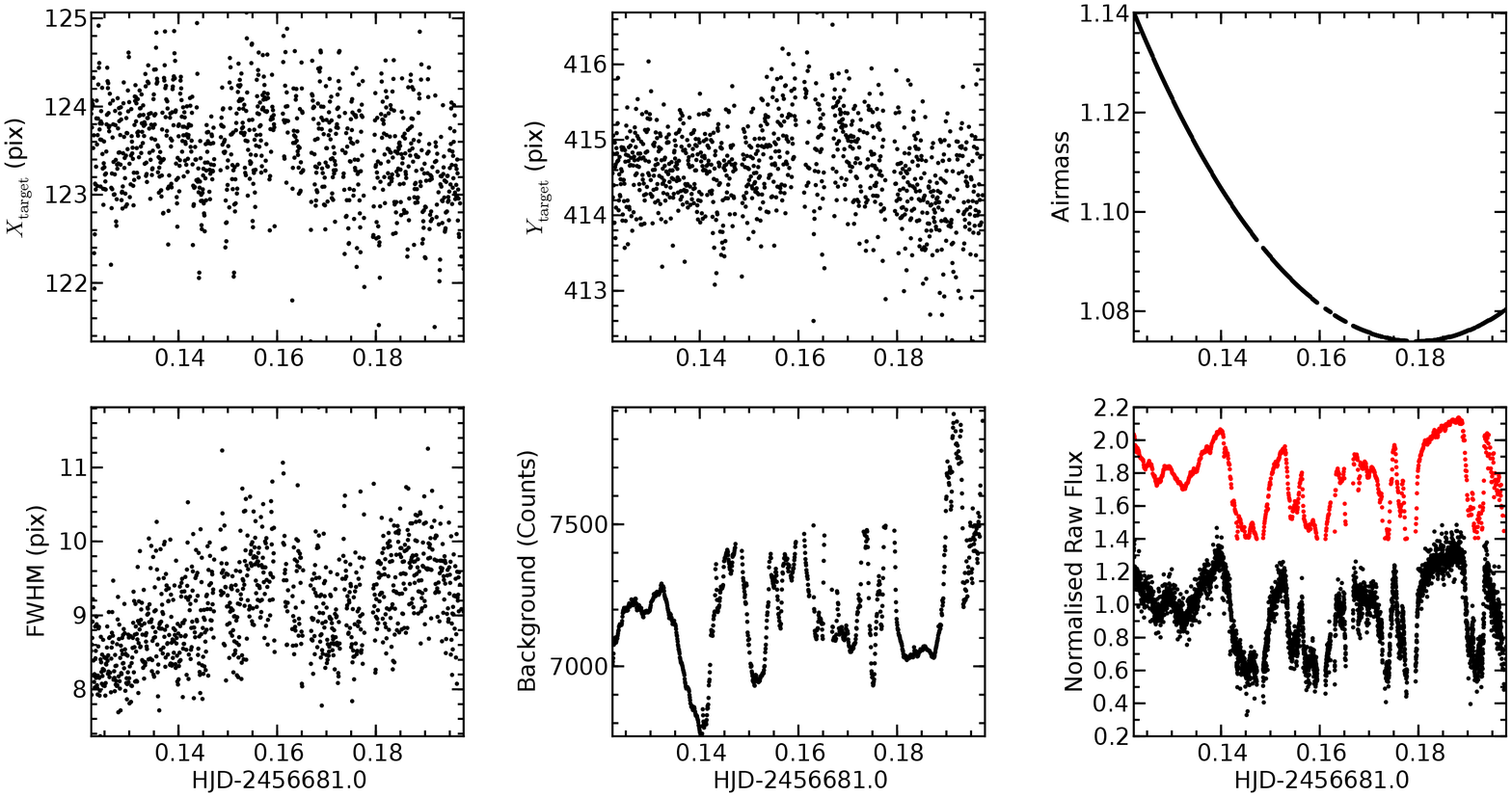}
  \caption{External parameters from the WASP-43b eclipse observation on 2014 January 23, including the target star detector $X$, $Y$ centroid positions, airmass, image PSF FWHM, sky background counts, and normalised raw fluxes for the ensemble of reference (black), and target (red, arbitrarily shifted upward by 0.8 for clarity) light curves.}  
\label{fig:W43_extern}
\end{figure*}

We observed the secondary eclipses of WASP-19b and WASP-43b using the IRIS2 instrument \citep{2004SPIE.5492..998T} on the 3.9\,m AAT at Siding Spring Observatory, Australia. IRIS2 uses a Hawaii 1-RG detector, read out over 4 quadrants, in double read mode. The instrument has a field of view of $7.7'\times7.7'$, and a pixel scale of $0.4486\,''/\text{pixel}$. The $Ks$ broadband filter was used for both observations, with bandwidth of $1.982 - 2.306\,\mu\text{m}$, centred at $2.144\,\mu\text{m}$.

The secondary eclipse of WASP-19b was observed on 2014 January 25 from 13:44--17:29 UT, consisting of 1009 frames with exposure times of 10 seconds per frame. A sample image of the WASP-19 field is shown in Figure~\ref{fig:field_img}. The telescope was guided throughout the eclipse sequence to ensure no drift over the $\sim 4$ hour observations. Dithered sequences before and after the observations were used to create a flat field image. The telescope was defocused to reduce the impact of imperfect flat fielding, inter- and intra-pixel variations on the photometry, as well as preventing saturation of the target and key reference stars, yielding stellar point-spread-functions (PSF) with full width half maxima (FWHM) of $\sim 4\,''$. Care was taken to ensure the target and key reference stars did not fall near bad pixels on the detector. Tests using dome flat fields showed the detector to be linear to within 1\,\% below 40000 counts. The target and reference star fluxes were kept well below this limit. The target remained above airmass 1.10, and the conditions were photometric throughout most of the eclipse sequence and the first dither sequence. The conditions were poorer during the final 1.2 hours of observations, however the final photometry was not severely affected and no frames from the eclipse sequence were rejected for the final analysis. Individual frames during the final dither sequence were interrupted by scattered clouds and were clipped out before they were combined to form the master flat field image. Figure~\ref{fig:W19_extern} shows the variation of the external parameters through the eclipse sequence, including the target position on the detector, airmass, FWHM, background counts, and raw flux of the target and reference stars. We noted that the target star drifted by only $\sim 1$ pixel through the $\sim 4$ hour eclipse sequence. 

The secondary eclipse of WASP-43b was observed on 2014 January 23 from 14:44--17:08 UT, consisting of 1440 frames with exposure times of 5 seconds per frame. A sample image of the WASP-43 field is shown in Figure~\ref{fig:field_img}. The observing strategy was the same as that described above for WASP-19b. The target remained above airmass 1.15 throughout the observations. The conditions were not ideal during the eclipse sequence, with frequent interruptions by scattered clouds. Object frames influenced by weather, identified as having target star raw fluxes $1\sigma$ below median, were removed before further analyses, leaving 973 measurements for analysis. Clouds also prevented a second set of dithered frames to be taken after the eclipse sequence. The external parameters from the WASP-43 observation are plotted in Figure~\ref{fig:W43_extern}, the target star drifted by less than 1 pixel over the observing sequence.

\subsection{Data reduction and light curve extraction}
\label{sec:data-reduction-light}

Dark subtraction, flat field correction, and bad pixel interpolation were performed for each object image.

Flat field images were created from the set of dithered frames, median combined after stars were masked. For the observations of WASP-19b, where dithered frames were taken before and after the eclipse observations, two separate flat field images were created. The two flat field images, $\text{Flat}_1$ and $\text{Flat}_2$, were combined with weights to create a master flat field image $\text{Flat}_\text{master,i}$ for every object frame $i$ according to:
\begin{equation}
  \label{eq:flat_weights}
 \text{Flat}_\text{master,i} = w_{1,i} \text{Flat}_1 + w_{2,i} \text{Flat}_2\,.
\end{equation}
The weights were generated such that the overall root-mean-square (RMS) of background of each object frame was minimised after flat field correction with $\text{Flat}_\text{master,i}$. Stars in each object frame were masked before the background RMS was calculated. The fitted coefficients of the linear combination of flat fields were highly correlated with the overall background count in an object frame, and varied smoothly through the eclipse sequence. We found the technique of combining the flats provides a first order correction for the temporal and spatial variations in the infrared sky background, and delivers more precise photometry than an equal weighted combination of the two flat fields. The technique is similar to that used by \citet{2014A&amp;A...563A..40C} for first order sky background corrections. The second set of dithered frames for the WASP-43 observation were not obtained due to clouds, so flat field division was performed using only the first set of dithered frames.

Aperture photometry for the target and reference stars were extracted using \emph{Source Extractor} \citep{1996A&amp;AS..117..393B}, with the coordinates matched and transformed to each frame using the \emph{FITSH} package \citep{2012MNRAS.421.1825P}. The aperture size is determined by the FWHM of stars in each image, and varies from image to image. We carried out tests with a variety of aperture sizes relative to the overall FWHM of each image, and found that an aperture of 2.5 $\times$ FWHM generated the most precise photometry for both the WASP-19b and WASP-43b target stars, as measured by the RMS of the pre-detrending out-of-eclipse light curves. By allowing the extraction aperture to vary as a function of the FWHM per image, rather than being fixed, we are accounting for the varying atmospheric conditions through the observations. The background beneath each target was subtracted by building a spatially interpolated background image via \emph{Source Extractor}. We find that subtraction of an interpolated background yielded light curves with lower RMS than the traditional background estimation over an annulus around each star. %We also tried PSF fitting photometry, but it yielded light curves with lower precision, by a factor of two, than our aperture photometry results.

For each set of observations, a master reference light curve was created from the ensemble of reference stars exhibiting stable photometry, with weights fitted so as to minimise the out-of-eclipse RMS scatter of the target light curve. In the absence of unstable reference stars, these weights converge to the flux ratio between the reference stars. To find the best photometric aperture for each reference star, we tested all permutations of the reference star -- aperture choices, choosing the permutation that yielded the least out-of-eclipse RMS for the target light curve. Seven reference stars were used in for the WASP-19 observation, and four reference stars were used for the WASP-43 observation. 

\subsection{Eclipse light curve fitting}
\label{sec:eclipse-fitting}

We fit the eclipse light curves using the \citet{1972ApJ...174..617N} models, with an adapted implementation of the \emph{JKTEBOP} code \citep{1981AJ.....86..102P,2004MNRAS.351.1277S}. The free parameters in our fit are the phase of the eclipse, determined by the orbital parameter $e\cos\omega$, and the depth of the eclipse, determined by the surface brightness ratio between the planet and the star $S_p/S_\star$. The system parameters period $P$, transit time $T_0$, planet to star radius ratio $R_p/R_\star$, normalised orbital radius $(R_p+R_\star)/a$, and line-of-sight inclination $i$ are taken from the most recent global analyses in literature: \citet{2013MNRAS.436....2M} for WASP-19 and \citet{2014A&amp;A...563A..40C} for WASP-43. We incorporated the uncertainties associated with each of these system parameters by drawing their values from Gaussian distributions with standard deviations as the literature uncertainty values. We convert the exposure time stamps to BJD-TDB using the \emph{UTC2BJD} tool \citep{2010PASP..122..935E}, such that they match with the literature $T_0$ values. The best fit values and uncertainties are explored via a Markov chain Monte Carlo (MCMC) analysis, using the \emph{emcee} ensemble sampler \citep{2013PASP..125..306F}. To better account for other error sources beyond photon-noise, we inflate the error bars for each point such that the reduced $\chi^2=1$ for the light curve before the start of the MCMC routine.  

Infrared time-series photometry are often heavily influenced by external parameters. For example, guiding errors can cause the $X$, $Y$ pixel position of the star on the detector to drift, leading to a systematic error in the resulting light curve. Changes to the FWHM will also change the pixels on which the stellar PSF falls, as well as affect the amount of stellar flux within the photometric aperture. Imperfect background subtraction can lead to correlations between the object light curve and the background counts. Our eclipse model $M$ is therefore a product of the theoretical eclipse light curve $M_\text{eclipse}$ and a model describing the influence of external parameters $M_\text{extern}$: 
\begin{equation}
  \label{eq:model}
  M = M_\text{eclipse}\left(t,e\cos\omega,S_p/S_\star \right)  M_\text{extern} \left( t,X,Y,F,B,A \right) \, .
\end{equation}
We incorporate into our model a linear correlation against a subset of the external parameters time $t$, detector positions $X$, $Y$, target FWHM $F$, background counts $B$, and target airmass $A$. Since the inclusion of more external parameters, and therefore more free parameters, will always lead to better fits to the data, we use the Bayesean Information Criterion (BIC) to determine the specific subset of external parameters that best model the light curve in each set of observations. We test subsets of these external parameters, each via its own MCMC minimisation, and adopt the analysis giving the minimum BIC for our final result.  We find for the WASP-19b eclipse,
\begin{equation}
  \label{eq:M_w19}
  M_\text{extern,W19} = c_0 + c_1 t + c_2 B + c_3 F \, ,
\end{equation}
and for the WASP-43b eclipse
\begin{equation}
  \label{eq:M_w43}
  M_\text{extern,W43} = c_0 + c_1 t + c_2 B + c_3 F \, ,
\end{equation}
where $B$ and $F$ are the variations of the background flux and target FWHM, respectively.

\begin{figure*}
  \centering
  \begin{tabular}{l}
     \textbf{WASP-19b}\\
    \includegraphics[width=18cm]{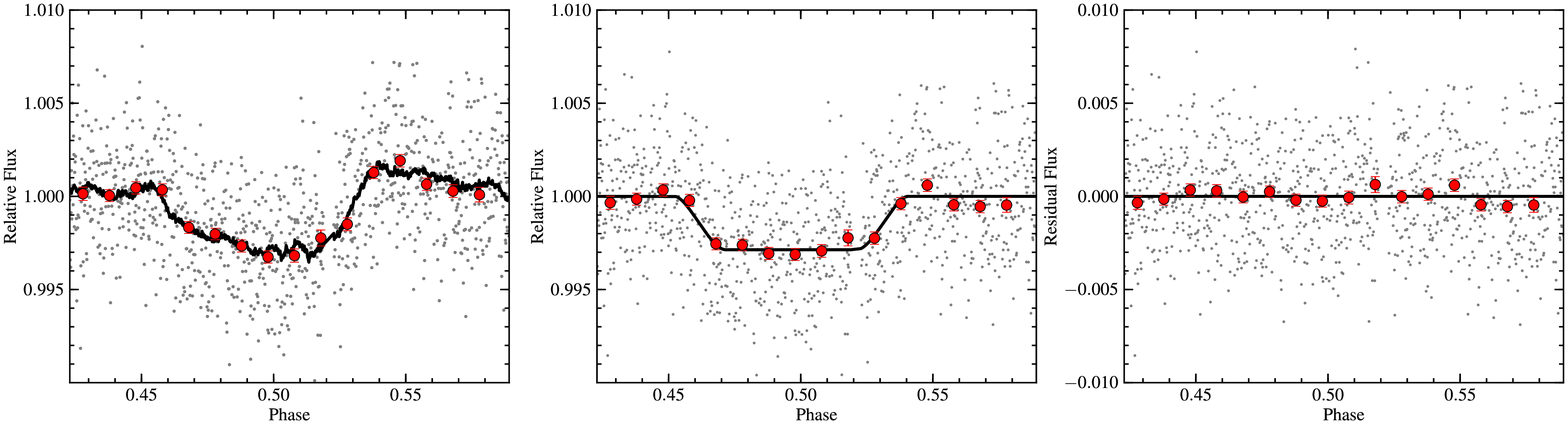}\\
    \textbf{WASP-43b}\\
    \includegraphics[width=18cm]{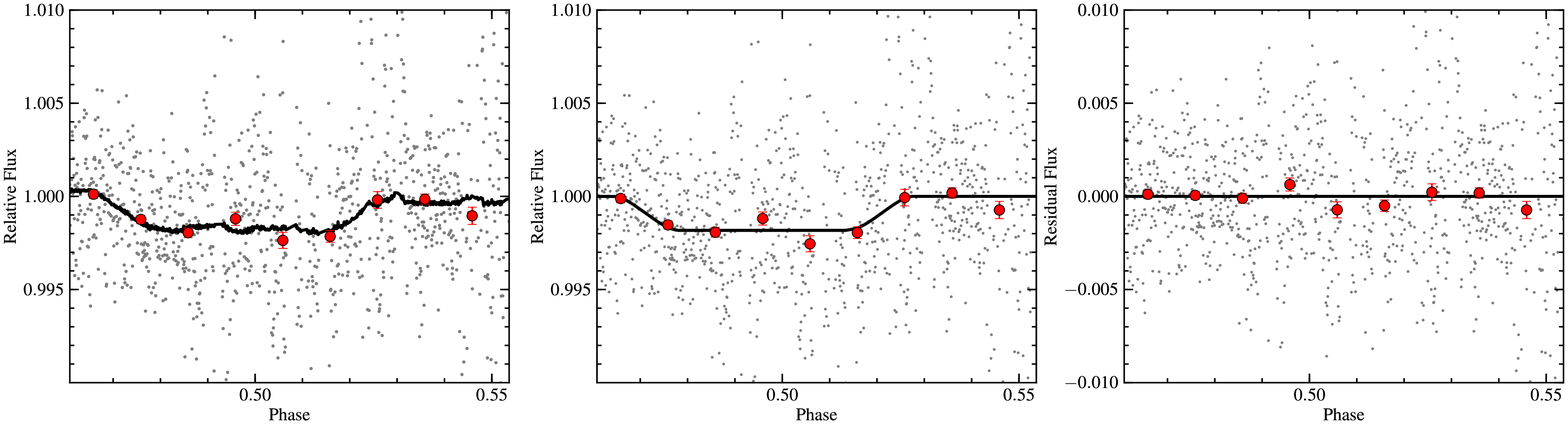}
  \end{tabular}

  \caption{The light curves for the eclipse of WASP-19b (\textbf{Top}) and WASP-43b (\textbf{Bottom}). \textbf{Left}: The pre-detrending light curves are plotted by the small points. Red points show the light curve binned at 0.01 in phase, with error bars derived from the scatter in the bin, assuming a poisson distribution. The black line shows the best fit eclipse model $M$, which incorporates both the theoretical eclipse light curve $M_\text{eclipse}$ and the influence of the external parameters $M_\text{extern}$ (see Equation~\ref{eq:model}). \textbf{Middle}: The eclipse light curve and best fit eclipse model, with the $M_\text{extern}$ component removed. \textbf{Right}: The light curve residuals, with the entire eclipse model $M$ removed.}
  \label{fig:lightcurve}
\end{figure*}

The eclipse light curves, models, and residuals, are plotted in Figure~\ref{fig:lightcurve}, and tabulated in Tables~\ref{tab:W19_lc} and \ref{tab:W43_lc}. We note that the precise tracking and guiding of the AAT minimised the influence of the external parameters, especially $X$ and $Y$, on the resulting light curves. 

\begin{table*} 
 \centering 
 \caption{WASP-19 Light curve} 
 \label{tab:W19_lc}
{\scriptsize \begin{tabular}{rrrrrrrrrrr} 
\hline \hline
BJD-TDB & Flux & Flux Error & Phase & $M$ & $M_\text{eclipse}$& X (pix) & Y (pix) & FWHM (pix) & Background (counts) & Airmass\\ 
\hline  
2456683.08460 & 0.99928 & 0.00139 & 0.42288 & 1.00033 & 1.00000 & 630.35500 & 324.49500 & 8.00000 & 11044.98000 & 1.09739\\ 
2456683.08473 & 0.99606 & 0.00139 & 0.42304 & 1.00035 & 1.00000 & 631.14400 & 326.04700 & 8.18000 & 11048.41000 & 1.09716\\ 
2456683.08583 & 1.00172 & 0.00138 & 0.42319 & 1.00035 & 1.00000 & 629.81700 & 325.73600 & 8.67500 & 11024.85000 & 1.09506\\ 
2456683.08595 & 0.99784 & 0.00138 & 0.42335 & 1.00041 & 1.00000 & 630.09100 & 325.66000 & 9.45500 & 11025.11000 & 1.09483\\ 
2456683.08608 & 1.00076 & 0.00139 & 0.42350 & 1.00016 & 1.00000 & 631.00900 & 325.53500 & 8.44000 & 11014.39000 & 1.09460\\ 
\hline
\end{tabular}} 
\begin{flushleft} 
This table is available in a machine-readable form in the online journal. A portion is shown here for guidance regarding its form and content.
\end{flushleft} 
\end{table*}

\begin{table*} 
 \centering 
 \caption{WASP-43 Light curve} 
 \label{tab:W43_lc}
{\scriptsize \begin{tabular}{rrrrrrrrrrr} 
\hline \hline
BJD-TDB & Flux & Flux Error & Phase & $M$ & $M_\text{eclipse}$& X (pix) & Y (pix) & FWHM (pix) & Background (counts) & Airmass\\ 
\hline 
2456681.12308 & 1.00190 & 0.00118 & 0.46086 & 1.00027 & 1.00000 & 123.91700 & 414.55800 & 8.52000 & 7006.94800 & 1.14013\\ 
2456681.12315 & 0.99971 & 0.00119 & 0.46094 & 1.00021 & 1.00000 & 123.97400 & 414.23800 & 7.92500 & 7016.24300 & 1.13998\\ 
2456681.12373 & 1.00330 & 0.00123 & 0.46102 & 1.00029 & 1.00000 & 121.93700 & 414.67600 & 8.20500 & 7084.41400 & 1.13857\\ 
2456681.12380 & 1.00239 & 0.00122 & 0.46110 & 1.00031 & 1.00000 & 122.55100 & 414.10200 & 8.02500 & 7083.70600 & 1.13842\\ 
2456681.12386 & 1.00006 & 0.00123 & 0.46118 & 1.00035 & 1.00000 & 123.12600 & 414.89100 & 8.85000 & 7088.28500 & 1.13827\\ 
\hline
\end{tabular}} 
\begin{flushleft} 
This table is available in a machine-readable form in the online journal. A portion is shown here for guidance regarding its form and content.
\end{flushleft} 
\end{table*}

\begin{figure*}
  \centering
  \begin{tabular}{ll}
     \textbf{WASP-19b}&    \textbf{WASP-43b}\\
    \includegraphics[width=7cm]{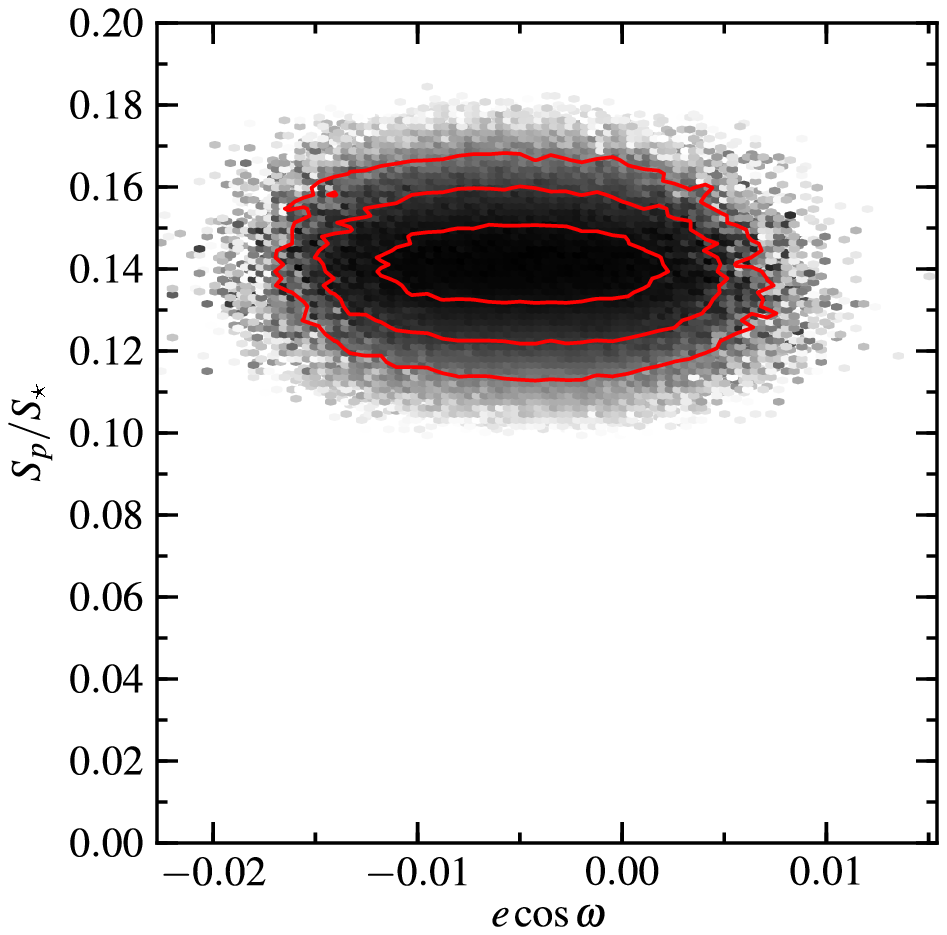}&
    \includegraphics[width=7cm]{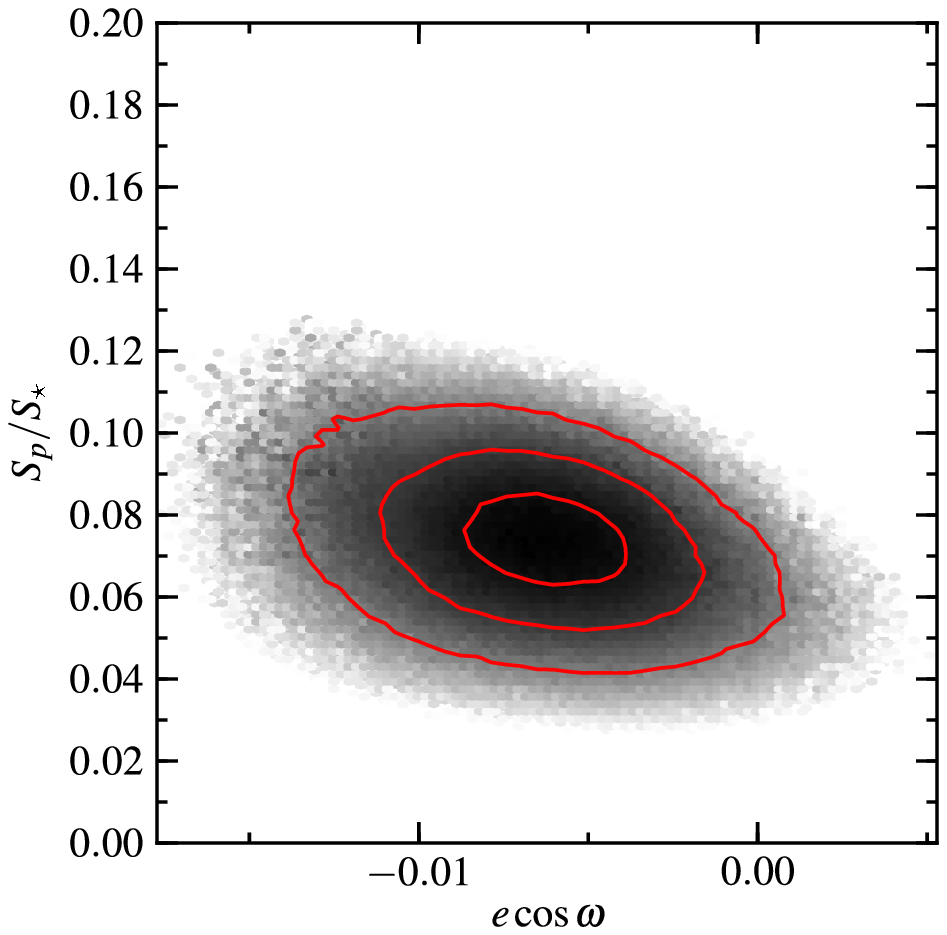}\\
  \end{tabular}
  \caption{Marginalised probability density plots showing the correlation between the eclipse phase, determined by $e\cos\omega$, and the eclipse depth, determined by the surface brightness ratio $S_p/S_\star$. The posterior for WASP-19b is plotted on the \textbf{left}, WASP-43b on the \textbf{right}. The 1, 2, and 3$\sigma$ confidence contours are marked by the red lines.}
  \label{fig:posterior}
\end{figure*}

Table~\ref{tab:eclipse_param} shows the results of the eclipse fitting analysis, including the orbit parameter $e\cos\omega$, and surface brightness ratio $S_p/S_\star$. The eclipse parameters are derived from the MCMC results, after marginalising over the parameters governing the external parameter correlation. We also include the derived eclipse depth and the $K_s$ band brightness temperature $T_B$. We assume black body spectra for the host stars, with stellar effective temperatures of $5460\pm90\,\text{K}$ for WASP-19 \citep{2013MNRAS.428.3164D}, and $T_\text{eff} = 4536_{-85}^{+98}\,\text{K}$ for WASP-43 \citep{2014A&amp;A...563A..40C}. The correlation between $e\cos\omega$ and $S_p/S_\star$ are shown by the marginalised probability density plots in Figure~\ref{fig:posterior}.

\begin{table}
  \centering
  \caption{Fitted and derived eclipse parameters}
  \label{tab:eclipse_param}

  \begin{tabular}{lrr}
    \hline \hline
    & \textbf{WASP-19b} & \textbf{WASP-43b}\\
    \hline
    \multicolumn{3}{l}{\textbf{Adopted Parameters$^{a}$}} \T\\
    $P$ (days) & $0.7888396 (10) $ & $0.81347437 (13)$ \\
    $T_0$ (BJD-TDB) & $2454775.33745 (35)$ & $2455934.792239 (40)$\\
    $R_p+R_\star/a$ & $0.33091(74)$ & $0.2325(20)$\\
    $Rp/R_\star$ & $0.14259(23)$ & $0.15739(41)$\\
    $i \, (^\circ)$ & $78.76(13)$ & $82.69(19)$ \\
    \multicolumn{3}{l}{\textbf{Free Parameters}} \T\\
    $e\cos\omega$ & $-0.0056_{-0.0057}^{+0.0070}$ & $-0.0062_{-0.0024} ^{+0.0022}$ \\
    $S_p/S_\star$ & $0.141_{-0.010}^{+0.010}$ & $0.074_{-0.011} ^{+0.011}$ \\
    \multicolumn{3}{l}{\textbf{Derived Parameters}} \T\\
    Depth (\%) & $0.287_{-0.020}^{+0.020}$ & $0.181_{-0.027}^{+0.027}$\\
   $T_B$\,(K) & $2310\pm60$ & $1743\pm67$ \B\\
   \hline
  \end{tabular}

\begin{flushleft} 
$^{a}${Adopted for the fitting routine. WASP-19b parameters from \citet{2013MNRAS.436....2M}, WASP-43b parameters from \citet{2014A&amp;A...563A..40C}. Uncertainties are given for the last two significant figures.}\\
\end{flushleft}

\end{table}

\begin{figure*}
  \centering
  \begin{tabular}{ll}
     \textbf{WASP-19b}&    \textbf{WASP-43b}\\
    \includegraphics[width=8cm]{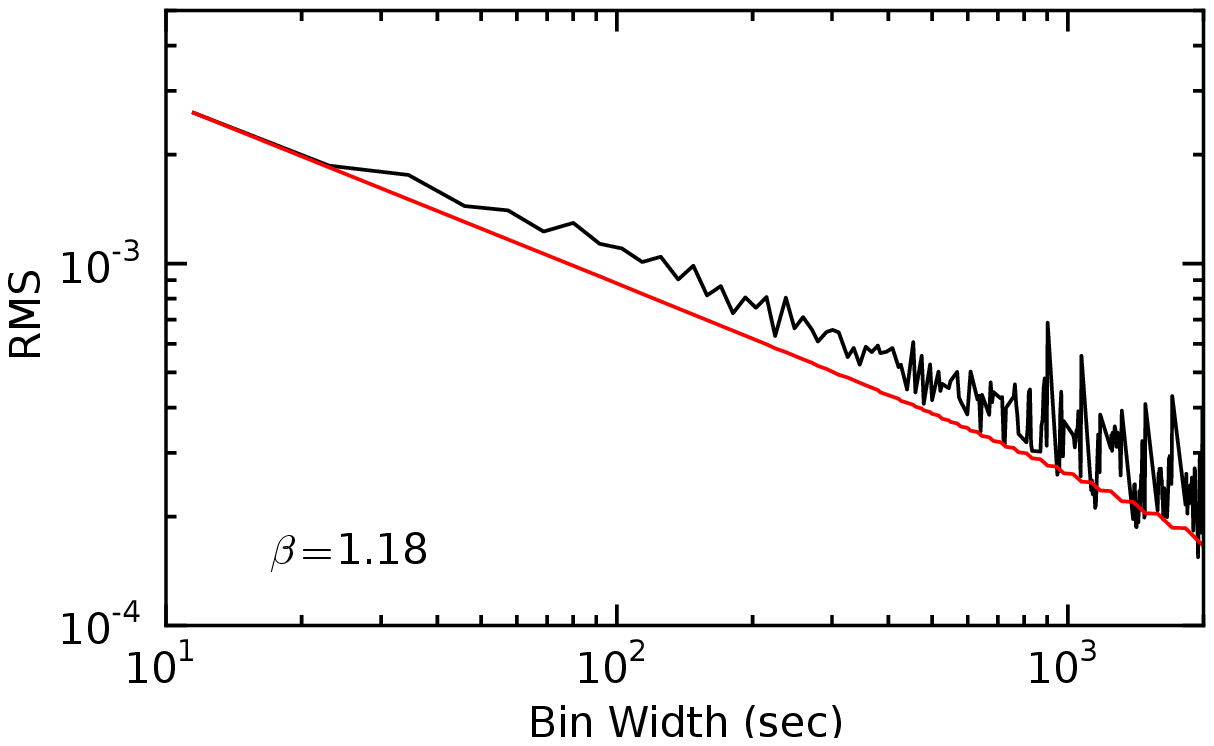}&
    \includegraphics[width=8cm]{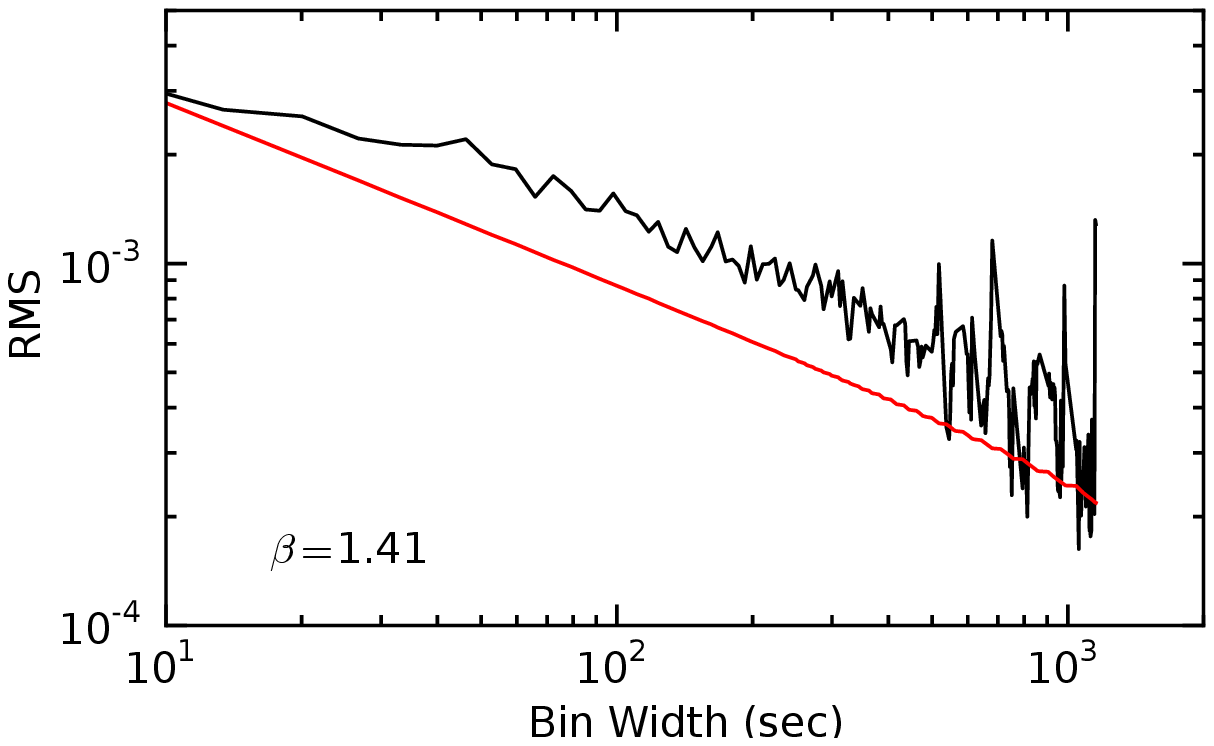}\\
  \end{tabular}

  \caption{$\beta$ factor diagnostic plots showing the amount of time correlated noise in the observations. The RMS of the light curve residuals as a function of the bin size for WASP-19b (\textbf{Left}) and WASP-43b (\textbf{Right}) are plotted in black. The expected $1/\sqrt{N}$ fall off for uncorrelated noise are plotted in red. The $\beta$ values for the light curves of each observation are also shown.}
  \label{fig:beta}
\end{figure*}

To examine for time-correlated noise that remains in the light curve residuals, we perform the $\beta$ factor diagnostic on each light curve as per \citet{2008ApJ...683.1076W}. We calculate the RMS of the residual light curves, with the model $M$ subtracted, binned into progressive larger bins. The RMS $\sigma_n$ of the light curve binned every $n$ points into $m$ bins, with respect to the RMS of the unbinned light curve $\sigma_1$ is given by 
\begin{equation}
  \label{eq:beta}
  \sigma_n = \beta \frac{\sigma_1}{\sqrt{n}} \sqrt{\frac{m}{m-1}} \, .
\end{equation}
For photon-limited data with no time-correlated noise, we expect $\beta=1$. Figure~\ref{fig:beta} plots the RMS of each light curve binned at different intervals. Time correlated signals exist in the residuals at the level of $\beta = 1.18$ for the WASP-19 observation, with an unbinned RMS scatter of 2.6\,mmag. The WASP-43 observation has an unbinned RMS scatter of 3.4\,mmag, with $\beta=1.41$, indicating a higher level of correlated noise. The expected scatter of the light curves is discussed further in Section~\ref{sec:iris2-detect-limits}.

\section{Discussion}
\label{sec:discussion}

\subsection{Comparison with previous observations}
\label{sec:results}

We observed the secondary eclipses of the irradiated hot-Jupiters WASP-19b and WASP-43b, measuring their eclipse depths to be $0.287_{-0.020}^{+0.020}$\,\% and $0.181_{-0.027}^{+0.027}$\,\% respectively. For WASP-19b, \citet{2010MNRAS.404L.114G} reported an eclipse depth of $0.366\pm0.072$\,\% in the 2.095\,$\mu$m narrowband filter, within $1.1\,\sigma$ of our result. Spectrophotometry by \citet{2013ApJ...771..108B} reported a depth of $0.227\pm0.016$\,\% in the 2.05-2.15\,$\mu$m wavelength bin, marginally consistent (within $2.3\,\sigma$) with our result. Averaging all the observations, the mean eclipse depth in $K$ band wavelength is 0.293\,\%, with standard deviation of 0.057\,\%. The mean error cited by the observations is 0.036\,\%, smaller than the standard deviation of the measurements by a factor of 1.6. \citet{2013ApJ...771..108B} calculated that stellar variability should only affect the eclipse depths at the $10^{-5}$ level, unable to account for the extra scatter in the three measurements. The $e\cos\omega$ measured from our observations is consistent within errors to the circular orbit measured by the \emph{Spitzer} secondary eclipses \citep{2013MNRAS.430.3422A}.

For WASP-43b, our measurement is consistent to within $1\sigma$ of \citet{2012A&amp;A...542A...4G} at 2.09\,$\mu$m ($0.156\pm0.014$\,\%), as well as being within $1\sigma$ of  \citet{2013ApJ...770...70W} and \citet{2014A&amp;A...563A..40C} in the $K_s$ band ($0.194\pm0.029$\,\% and $0.197\pm0.042$\,\% respectively). The mean and standard deviation of the eclipse depths measured by the four observations is $0.182\pm0.016$\,\%, with a mean quoted error of $0.028$\,\%. In this case, the quoted errors represent well the scatter in the repeat observations. The $e\cos\omega$ measured from our observations is marginally inconsistent, at $2.6 \sigma$, with the expectation of a circular orbit and the phase of the eclipse measured by \emph{Spitzer} \citep{2014ApJ...781..116B}. This inconsistency may be due to the higher red noise from the poorer weather conditions of the observation. Uncorrected correlated systematics may also cause the weak correlation seen in the posterior distribution (Figure~\ref{fig:posterior}) between $e\cos\omega$ and $S_p/S_\star$.

Demonstrating the repeatability of eclipse measurements is key to providing robust constraints on the physical properties of the hot-Jupiters. Using hot-Jupiters with multiple \emph{Spitzer} secondary eclipse observations in the same band, \citet{2014arXiv1402.6699H} showed that the average reported observational uncertainties of \emph{Spitzer} observations are underestimated by a factor of 2. Repeat ground-based secondary eclipse observations are not common. In addition to WASP-19b and WASP-43b, only WASP-12b has three or more ground-based observations in the same photometric band \citep{2011AJ....141...30C,2012ApJ...744..122Z,2012ApJ...746...46C,2012ApJ...760..140C}, and three others have single repeat observations \citep[CoRoT-1b, TrES-3b, WASP-33b;][]{2009ApJ...707.1707R,2012ApJ...744..122Z,2009A&amp;A...493L..35D,2010ApJ...718..920C,2012ApJ...754..106D,2013A&amp;A...550A..54D}. The mean scatter in the measured eclipse depth between repeat observations is $1.4 \times$ larger than the mean uncertainty estimate of the observations. Some analysis techniques have been developed to yield more realistic parameter uncertainties, such as the widely adopted practice of inflating light curve point uncertainties by the reduced \chisq or $\beta$ values, or approximating correlated noise using wavelets \citep[e.g.][]{2009ApJ...704...51C} and Gaussian processes \citep[e.g.][]{2012MNRAS.419.2683G}. However, repeat observations are the best approach for providing robust uncertainties when interpreting atmosphere model retrievals.

\subsection{VSTAR atmosphere models }
\label{sec:vstar-atmosph-models}

We used the line-by-line radiative transfer code, Versatile Software for Transfer of Atmospheric Radiation (VSTAR) \citep{2012MNRAS.419.1913B} to derive models for the atmospheres of both planets, WASP-19b and WASP-43b. We applied $\chi^{2}$ minimisation to derive the best fitting models to data previously published and presented here.
 
Currently two different modelling techniques are exploited to obtain atmospheric composition of the extrasolar planets. The first one is a direct method that assumes the atmospheric temperature and pressure distribution as a function of height based on the self-consistent solution that conserves the net flux from the planetary atmosphere \citep{2005ApJ...627L..69F}. The radiative transfer calculation is then performed with a desired spectral resolution. The other method is the forward retrieval modelling of the atmospheric parameters that uses available intensity data at different wavelengths to provide the posterior probability of the temperature versus pressure profile and atmospheric chemical abundances. Forward retrieval models give best results for the datasets that are well sampled over large span of wavelengths. Currently such datasets are available only for a handful of extrasolar planets. 

We note that a treatment of the atmospheric composition of hot Jupiters within one dimensional domain is a striking oversimplification. The intense irradiation of nearby and most likely tidally locked planets leads to dynamically complex atmosphere with strong winds due to the heat redistribution from the side facing a star and with the regions that divert from assumption of thermodynamics equilibrium \citep[e.g.][]{2002A&amp;A...385..166S,2003ApJ...587L.117C,2009ApJ...699..564S,2012ApJ...750...96R}. Strong temperature differences between day and night sides lead to differences in pressure versus temperature (P-T) profiles at different positions on the planet that has an effect on atmospheric composition and cloud formation. A proper understanding of such conditions requires a combined modelling of both transit transmission spectroscopy, phase variations, and eclipse emission spectroscopy data using 3D global circulation models \citep[e.g.][]{2013MNRAS.435.3159D}.

However given the sparse amount of data available for both planets, we decided to use a direct modelling technique. The single P-T profile that approximates atmospheric conditions on the day side for each planet that is probed in our measurements was adapted from previously published direct \citep{2014ApJ...781..116B,2012ApJ...758...36M} and forward retrieval models \citep{2014ApJ...783...70L}. 

Although it is difficult to assert the uniqueness of the best fitting spectral models, currently the P-T profiles derived from the forward retrieval scheme by \citet{2014ApJ...783...70L} best reproduce the observations. Admittedly these parameters are derived for atmospheres with much simpler compositions than assumed in our direct models presented here.

In the first step of our VSTAR modelling, we create a multilayered atmospheric structure derived from thermochemical equilibrium calculations, with the assumed metallicity and the carbon to oxygen (C/O) ratio, using the Ionization and Chemical Equilibrium (ICE) package of VSTAR. The equilibrium abundances of chemical ingredients are expressed in terms of mixing ratios, which are obtained for known pressures and temperatures in every atmospheric layer using a database of 143 compounds in gaseous and condensed phases. We derive mixing ratios, in chemical equilibrium, for the following molecules and atomic species: H$_2$O, CO, CH$_4$, CO$_2$, C$_2$H$_2$, HCN, TiO, VO, Na, K, H$_2$, He, Rb, Cs, CaH, CrH, MgH and FeH that are thought to be relevant in the atmospheres of the highly irradiated "hot" Jupiters. We note that disequilibrium effects, such as photochemistry, will alter the chemistry of the upper atmosphere from these models \citep[e.g.][]{2011ApJ...737...15M,2012ApJ...745...77K}. 

In the subsequent step multiple-scattering radiative transfer calculations are performed for every wavelength on the grid with a specified spectral resolution. The opacities in this final spectrum are derived by using a comprehensive database of molecular spectral lines described in \citet{2012MNRAS.419.1913B}. In addition we include Rayleigh scattering by H$_2$, He,  H, the opacities from the  collisionally induced absorption (CIA) due to H$_2$--H$_2$ and H$_2$--He \citep{1998STIN...9953342B} and the free-free and bound-free absorption from H, H$^{\_}$ and H$_2$$^\_$.  We listed all molecular and atomic absorbers used in our models with references to the line databases in Table 2 of  \citet{2013ApJ...774..118Z}. The spectra of host stars WASP-19 and WASP-43 were obtained from the STScI stellar atmosphere models \citep{2004astro.ph..5087C} for stars with effective temperature, T$_\text{eff}$ of 5500K and 4500K respectively. Finally, we explore the effect of clouds and hazes models to the model spectra. The haze model assumes particles with similar properties to enstatite, and assume spherical symmetry with absorption and scattering calculated using Mie theory \citep{2002sael.book.....M}.

\subsubsection{The atmosphere of WASP-19b}
\label{sec:atm-wasp-19b}

\begin{figure*}
  \centering
 \includegraphics[width=13cm]{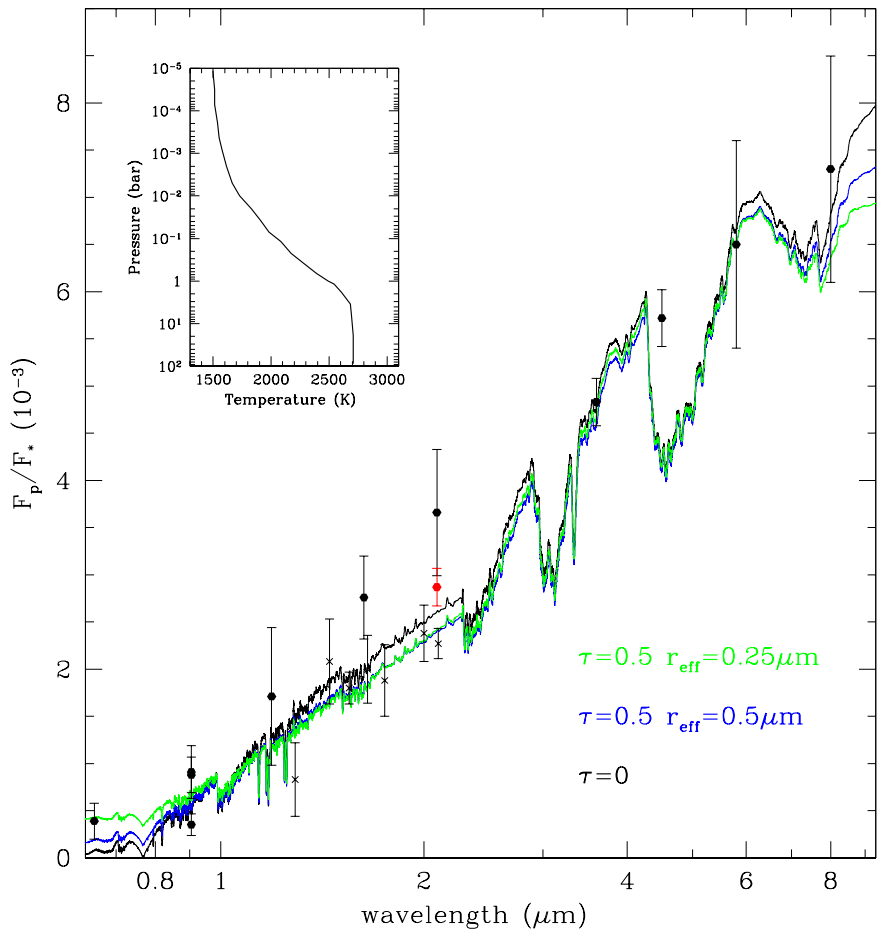}
  \caption{The effect of a top-layer haze on the modelled emission spectrum of WASP-19b. VSTAR model spectra of WASP-19b derived from the P-T profile in the inlay, with C/O ratio of 1.0. The cloudless model (in black) is over plotted with the model with haze at the top of the atmosphere with optical depth of $\tau =0.5$ at 1$\mu$m and  mean particle size of 0.5 $\mu$m (in green) and 0.25 $\mu$m (in blue). The IRIS2 $K_s$ band eclipse is plotted in red. Previous photometric eclipse measurements are marked in black. Spectrophotometric measurements from \citet{2013ApJ...771..108B} are marked by the grey crosses.  }
  \label{fig:W19_spec1}
\end{figure*}

The application of VSTAR models to WASP-19b was described in \citet{2013ApJ...774..118Z}, where we explored the effect of changing C/O abundance, as well as the addition of hazes in the top layers of its atmosphere. In particular, we suggested that WASP-19b hosts a carbon-rich atmosphere, that was also consistent with modelling by \citet{2012ApJ...758...36M}. This high C/O ratio is mainly supported by the high flux ratio exhibited by the planet at $<1\,\mu\text{m}$. Perhaps not surprisingly a forward retrieval study by \citet{2014ApJ...783...70L}, based on the broadband infrared ground and \emph{Spitzer} observations, was not successful in constraining C/O abundance in WASP-19b. \citet{2014ApJ...783...70L} did not include the five optical $(<1\,\mu \text{m})$ measurements, nor the spectrophotometric eclipse observation by \citet{2013ApJ...771..108B}, in their analysis. 

Previously we found it difficult to obtain a single P-T profile that resulted with the model that fits consistently all published data, especially given significant differences between NIR data points from \citet{2013ApJ...771..108B} and \citet{2013MNRAS.430.3422A}. Here we test models for a dense grid of the P-T profiles that range between our 'hotter' and 'cooler' profiles from \citet{2013ApJ...774..118Z} that are calculated for C/O corresponding to the solar abundance and the enhanced C/O=1. The tested P-T profiles had the same shape as in \citet{2012ApJ...758...36M} but were displaced from it by an addition of constant temperature offsets to all its points. In the iterative process of minimising $\chi^2$ for models in the grid and exploring smaller offsets around the best fitting models we obtained marginally better fits for the models with C/O=1, with a best $\chi^2$ value of 46.9 for the model derived from the P-T profile shown in Figure~\ref{fig:W19_spec1} that is slightly cooler than the profile used in \citet{2012ApJ...758...36M}.

Rayleigh scattering due to the presence of hazes and clouds in the upper atmosphere can also lead to enhanced flux ratio at the optical to NIR range of wavelengths, that seems to explain well the nearly-featureless transmission spectrum in this region for HD189733b \citep{2011MNRAS.416.1443S,2008MNRAS.385..109P}. A recent measurement of albedo in the secondary eclipse spectrum of HD189733b \citep{2013ApJ...772L..16E} with the STIS on the Hubble Telescope seems to support the haze hypothesis. However a detailed composition and distribution of hazes in atmospheres of the hottest Jupiters is still unclear. Highly refractory compounds such as perovskite, corundum or more abundant enstatite suggested as the component of haze in HD189733b, form condensates in temperatures lower than $\sim$2000~K at the top of WASP-19b. In our model a hypothetical haze characterised by the same optical properties as enstatite was placed in the top layer of the atmosphere and we varied its optical depth and the mean size of its particles. In \citet{2013ApJ...774..118Z} we found that models with particles smaller than 0.5 $\mu$m tend to overestimate the flux reflected in the optical part of the spectrum. A model with haze with optical depth of $\tau =0.5$ at 1$\mu$m at the top of the atmosphere composed of particles with mean size of 0.5 $\mu$m and 0.25 $\mu$m is shown in Figure~\ref{fig:W19_spec1}. These two models fit the data only marginally better than our best fitting cloudless model. However if hazes or clouds define the features of the optical spectrum, they cannot be optically thick in the near infrared where the recent HST observations of transmission spectrum of WASP-19b by \citet{2013MNRAS.434.3252H} show a possible detection of water absorption. We caution that the significance of the preference for haze models depends heavily on the value and uncertainty of the optical measurements, and repeated observations in that wavelength regime are require to test the models presented.

\subsubsection{The atmosphere of WASP-43b}
\label{sec:atm-wasp-43b}

\begin{figure*}
  \centering
 \includegraphics[width=12cm]{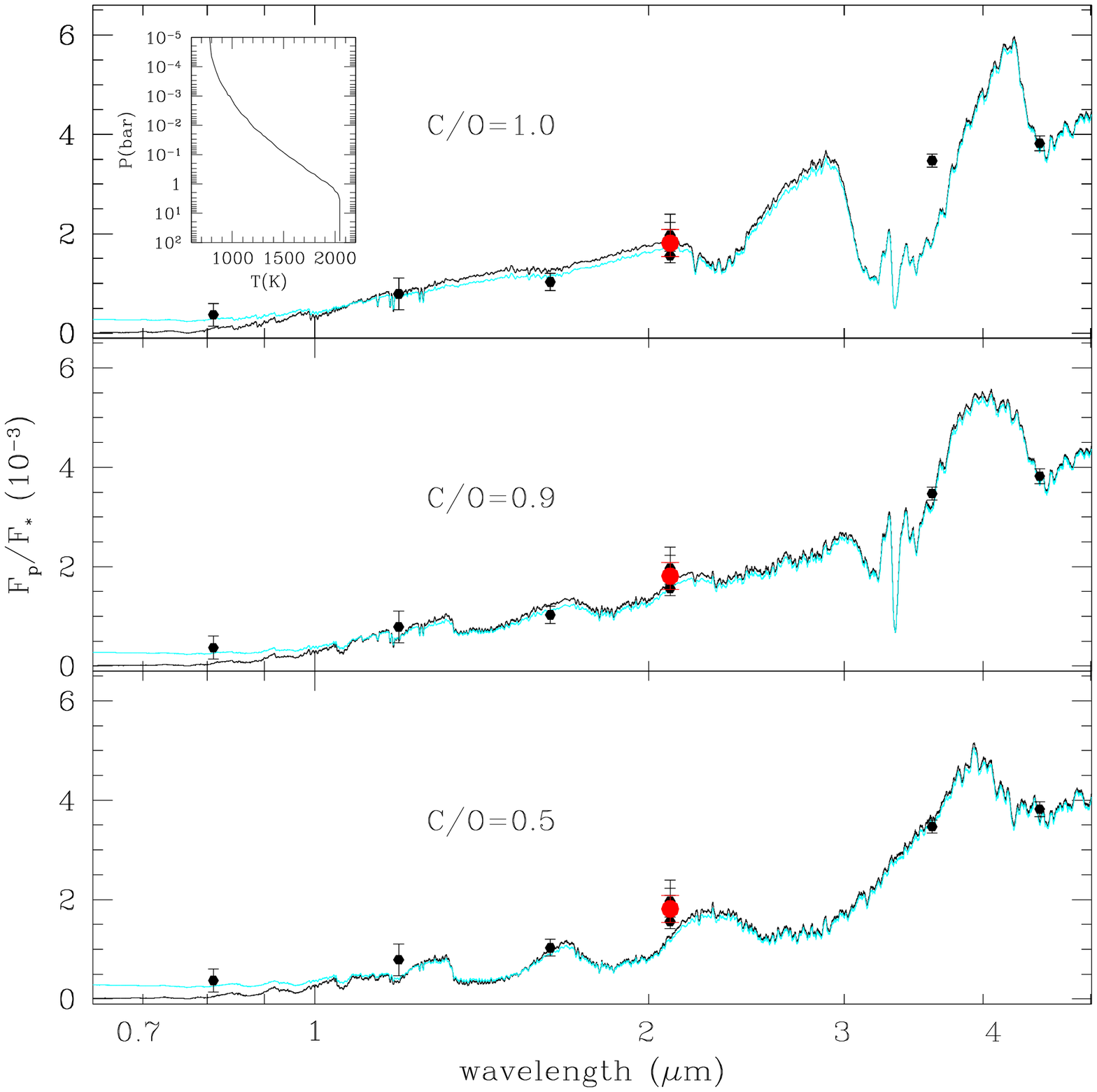}
  \caption{The effect of C/O abundance ratio on the modelled emission spectrum of WASP-43b. VSTAR model spectra of WASP-43b derived from the P-T profile in \citet{2014ApJ...781..116B} with  solar metallicity, with C/O ratios between 0.5 and 1.0. The cloudless models (black lines) are over plotted with the models with haze at the top of the atmosphere with optical depth of $\tau =0.5$ at 1$\mu$m and mean particle size of 0.25 $\mu$m (thin lines). The literature measurements are plotted in black, IRIS2 measurement in red.} 
  \label{fig:W43_spec}
\end{figure*}

\begin{figure}
  \centering
 \includegraphics[width=9cm]{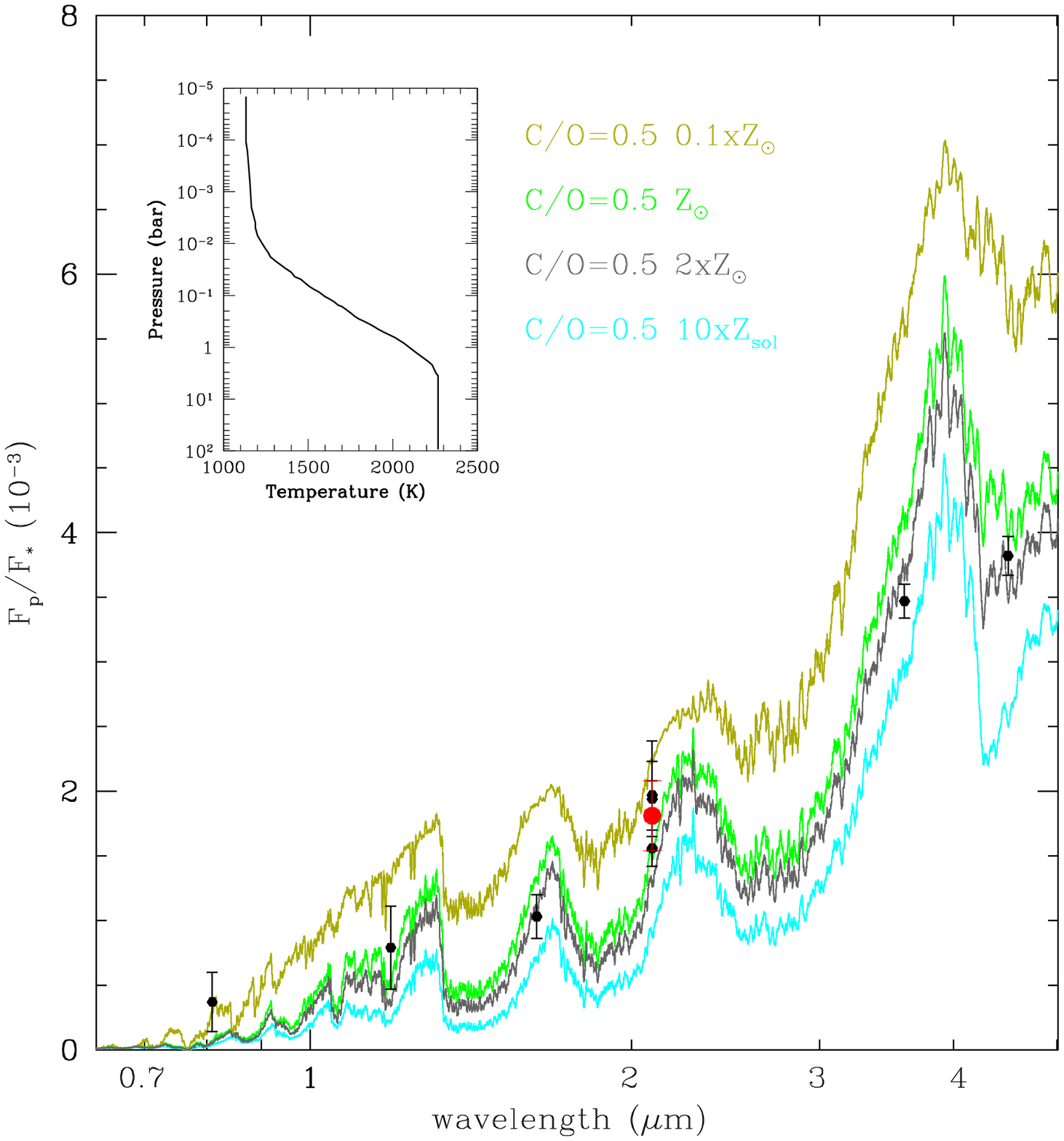}
  \caption{The effect of metallicity on the modelled emission spectrum of WASP-43b. VSTAR model spectra of WASP-43b derived from the hotter P-T profile in \citet{2014ApJ...781..116B} for four different metallicities assuming C/O = 0.5. The literature measurements are plotted in black, IRIS2 measurement in red. } 
  \label{fig:W43_spec1}
\end{figure}

Although WASP-43b shows relatively deep secondary eclipses due to the large planet to star radius ratio, making it a good target even for the ground telescopes, currently available data from studies listed in Section~\ref{sec:introduction} cover only sparsely the optical and infrared wavelength range. In two studies, \citet{2014ApJ...781..116B} and \citet{2014ApJ...783...70L} provided models of  the planetary spectrum using two different methods: a direct retrieval based on the assumed physical structure of the atmosphere and the forward retrieval of atmospheric parameters, such as the P-T profiles and mixing ratios of the few selected molecules, respectively. Both methods are consistent with the lack of temperature inversion in the atmosphere of WASP-43b.  

In the VSTAR modelling of the WASP-43b spectrum we tested the P-T profiles used by  \citet{2014ApJ...781..116B} for solar (Z$_\odot$) and 10 times higher than solar metallicity (10xZ$_\odot$). We also used the best-fit profile obtained by \citet{2014ApJ...783...70L} that indicated a significantly hotter upper atmosphere of the planet. We computed model spectra using the same sources of opacity as for WASP-19b (Section~\ref{sec:vstar-atmosph-models}). For each P-T profile we constructed models with C/O ratios between 0.5 and 1.0. We also tested atmospheres with different metallicities for C/O ratio of 0.5 for the hotter P-T profile from \citet{2014ApJ...781..116B}, because we found that models with highly enhanced metallicity tend to underestimate the level of flux as compared with the measurements from \emph{Spitzer} (Figure~\ref{fig:W43_spec}). The best agreement with the \emph{Spitzer} data was shown by our models with about twice Z$_\odot$ (Figure~\ref{fig:W43_spec1}). One should note that some discrepancies between our models and those published previously could be accounted for if the line databases used in different modelling schemes were not uniform. 

Models with C/O ratio of 0.5 for all tested profiles tend to underestimate the K-band data. In Figure~\ref{fig:W43_spec}  we plot the cooler P-T profile from \citet{2014ApJ...781..116B}, obtained assuming solar metallicity for different values of C/O. We find a best fit C/O ratio of 0.9 for the models.

Nevertheless it is clear that none of  these models can reproduce the level of reflectivity of the atmosphere in the optical that is implied from the measurement in i band \citep{2014A&amp;A...563A..40C}. As in the models of WASP-19b, we followed on to derive the spectra of WASP-43b in the presence of haze with varied optical depth at the top layer of the atmosphere. We find that the models with P-T profile from \citet{2014ApJ...783...70L} seem to lead to a better agreement with the data, provided the C/O ratio is lower than 0.7. On the other hand the `hazy' models obtained by the cooler P-T profile from \citet{2014ApJ...781..116B} again fits data best at C/O = 0.9. After testing different values of optical depth required to fit the i-band point we found the lowest $\chi^2$ value of 4.5 for the model with $\tau =0.5$ at 1$\mu$m and mean particle size of 0.25 $\mu$m. Smaller particles and higher optical depths of the haze tend to increase reflectivity of the atmosphere, inconsistent with the measurement by \citet{2014A&amp;A...563A..40C}. Clearly more data obtained with better precision are needed throughout optical spectrum in order to provide constrains sufficient to model detailed properties of the haze in the atmosphere of this planet.  

In Figure \ref{fig:W43_spec} the models with low C/O ratio show strong water absorption bands in near infrared that entirely disappear in models with C/O$\geq1$. These features could be helpful in differentiating between carbon rich and poor atmospheres. Unfortunately for ground-based observations these bands overlap with the water vapour absorption in the earth atmosphere. A new technique that we proposed in \citet{2014MNRAS.439..387C} relies on modelling and removing telluric features from the ground-based spectroscopic data. It provides better results than using a traditional telluric standard stars for this purpose, which will make observations in these spectral windows more reliable. In the far infrared, the spectrum of WASP-43b is dominated by CO$_2$ and CO bands for the oxygen-rich atmospheres while the influence of CH$_4$, HCN and C$_2$H$_2$ absorption, as well as stronger CO features, are observed for higher C/O ratios. Our assertion about the C/O abundance weighs heavily upon just two data points from the \emph{Spitzer} observations, therefore is should be treated with caution. However testing such an array of parameters in these models is useful for understanding  which targeted observations could potentially constrain the composition of WASP-43b atmosphere in the future.

\subsection{AAT+IRIS2 Detection Limits}
\label{sec:iris2-detect-limits}

This paper presents the first high precision photometric time series study performed using the AAT+IRIS2 facility. Here, we perform a set of signal injection and recovery exercises to explore the capabilities of the instrument and its potential for future work in eclipse and transit observations. We explore the detectability of an eclipse as a function of its depth and the brightness of the target star. 

\begin{figure}
%  \centering
  \includegraphics[width=8cm]{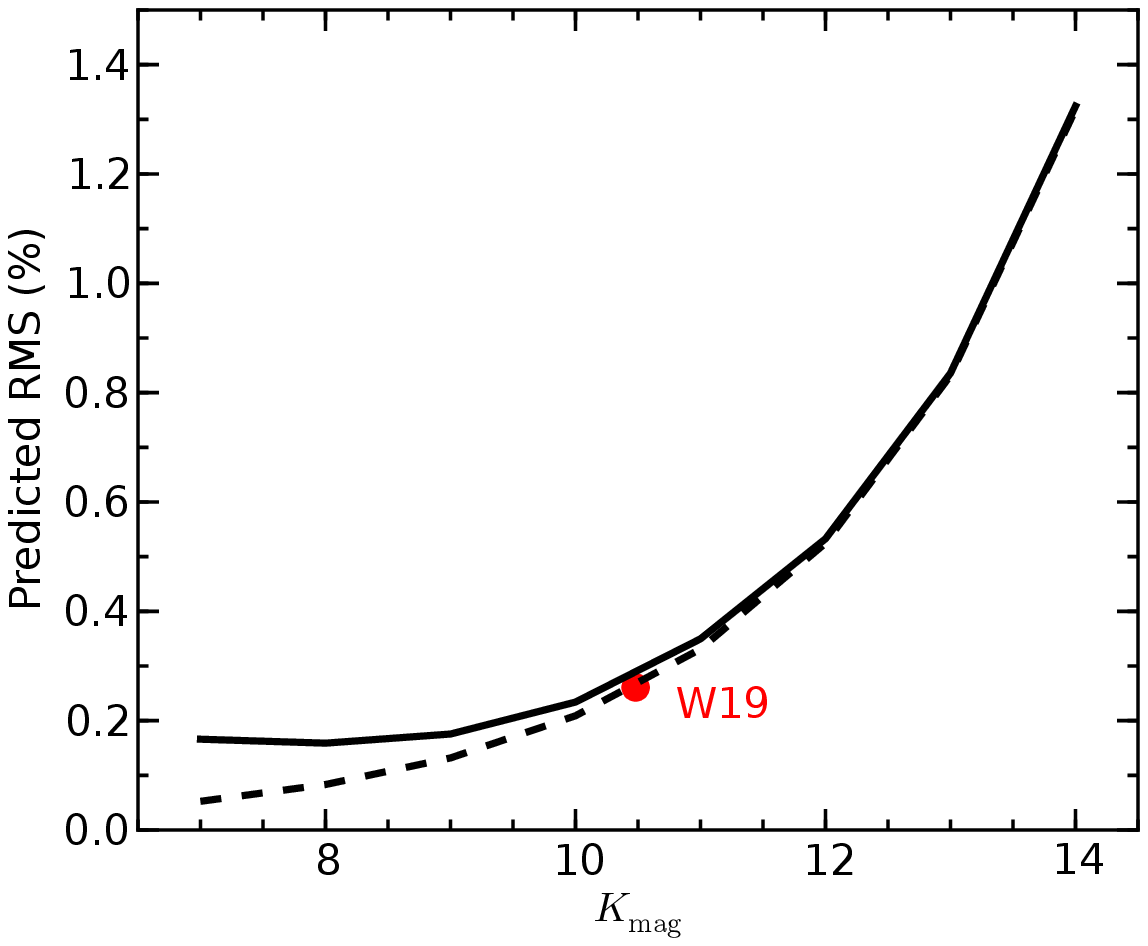}
  \includegraphics[width=9cm]{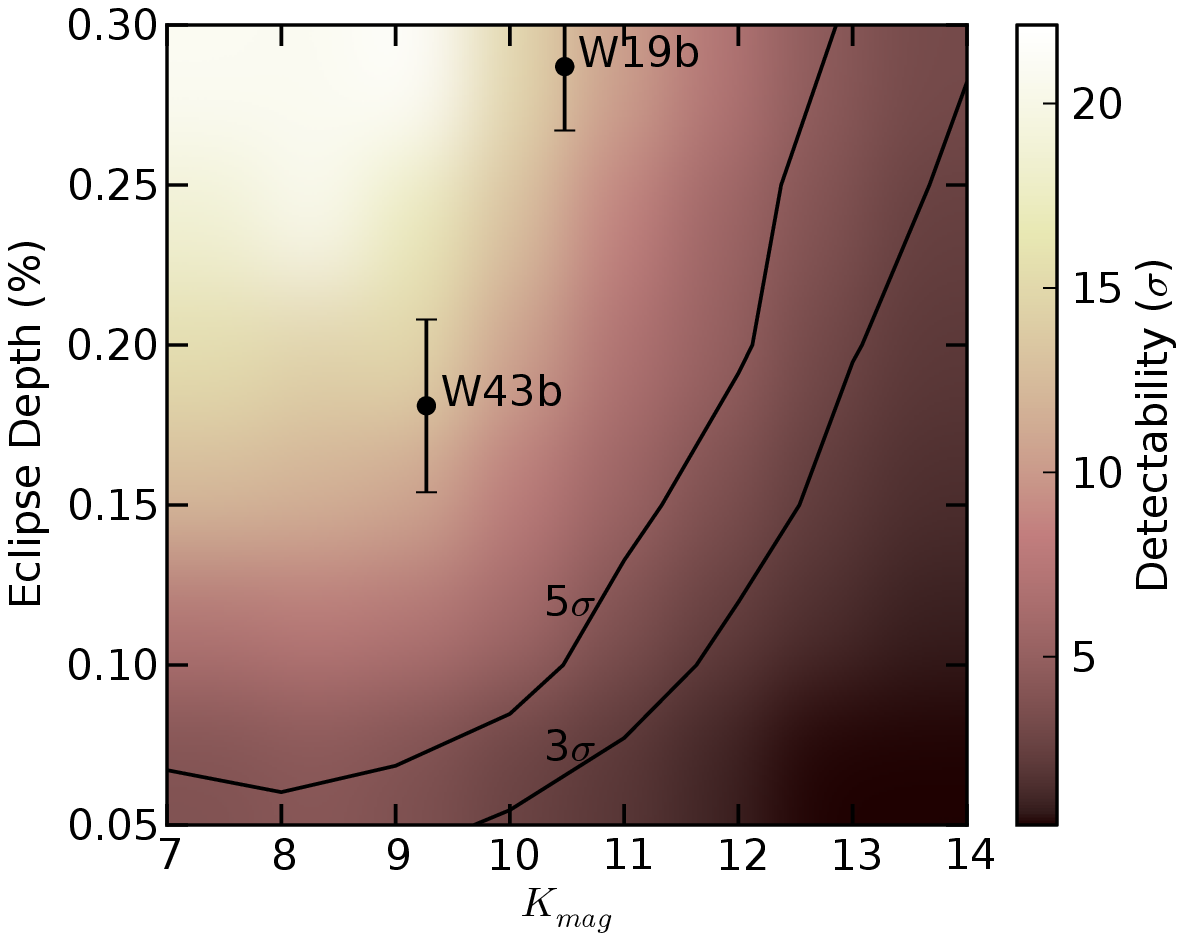}
  \caption{The expected photometric precision and eclipse detectability of the AAT+IRIS2 system. \textbf{Top:} The solid line marks the predicted RMS for an IRIS2 light curve, at 10s cadence, with respect to the $K_\text{mag}$ of the target. The precision obtained from our WASP-19 observation is marked by the red point. The dashed line representes the expected precision, scaled from the WASP-19 observation, if only photon count is accounted for. \textbf{Bottom:} The AAT+IRIS2 detectability for the eclipse signal of WASP-19b as a function the target star magnitude and eclipse depth, as derived from signal injection and recovery. The 3 and $5\,\sigma$ detection thresholds are marked by the contours. Our WASP-19b and WASP-43b measurements are plotted. }
  \label{fig:detectability}
\end{figure}

To characterise the dependence of the photometric precision as a function of the number and brightness of the available reference stars, we extract the light curves of 17 stars in the WASP-19 observation, with brightnesses between 0.1 and 3 times that of WASP-19. We then calculated the RMS of the residual WASP-19 light curves corrected against all possible 131071 subset combinations of the 17 reference stars. To generalise the dependence, we interpolate the measured RMS as a function of the number of reference stars in a subset $N$, and the ratio between the flux of the target and the sum flux of that subset of reference stars $F_\text{ratio}$\,:
\begin{equation}
  \label{eq:fratio}
  F_\text{ratio} = \frac{ \sum_i^N F_{\text{ref},i} } {F_\text{target}} \,. 
\end{equation}
We find a primary dependence on $F_\text{ratio}$, and a weaker inverse dependence on $N$. That is, fewer brighter reference stars produce higher precision photometry than numerous fainter reference stars, given the same sum reference flux. We then generalise for target stars of any brightness in any field via a 2D interpolation over $F_\text{ratio}$  and $N$. For an average field, the mean density of stars as a function of brightness is determined from the 2MASS catalogue, averaged over the entire sky. For a target star of some magnitude $K_\text{mag}$, we assume all stars within the IRIS2 field of view within magnitudes $K_\text{mag}$ and $K_\text{mag}+2$ are viable reference stars, and determine the associated $F_\text{ratio}$ and $N$. Field stars brighter than the target are often saturated in an observation, and therefore discarded. The expected photometric precision as a function the target star is plotted in Figure~\ref{fig:detectability}. 

We check the consistency of this interpolation with the WASP-43 observation. The WASP-43 light curve has a RMS scatter of 3.4\,mmag at the 5s cadence of the observation, equivalent to 2.4\,mmag at 10s cadence. If we simply scale the RMS scatter of the WASP-19 observation (2.6\,mmag) by the brightness of the targets ($K_\text{mag}=9.3$ for WASP-43, 10.5 for WASP-19), the expected 10s cadence RMS scatter of the WASP-43 observation should be 1.5\,mmag. However, the WASP-43 field has significantly fainter reference stars, with $F_\text{ratio} = 1.2$ and $N=4$, compared to $F_\text{ratio}=12$ and $N=7$ for the WASP-19 field. Using the generalisation above, we calculate an expected 10s cadence RMS scatter for WASP-43 of 2.4\,mmag, agreeing with the actual observation.

The eclipse signals are injected into the WASP-19 light curve residuals, after the eclipse signal $(M_\text{eclipse})$ is subtracted. To best simulate the observed light curves, we do not remove the external parameter correlated signals $(M_\text{extern})$. A grid of simulated light curves, with target $K_\text{mag}$ from 7 to 14 at steps of 1 mag, and eclipse depth from 0.05\% to 0.30\% at steps of 0.05\%, were created. The other system parameters of the injected eclipse (period, $R_p/R_\star$, $(R_p + R_\star)/a$, and $i$) are identical to that of WASP-19b. The eclipses are then recovered using the MCMC analysis as per Section~\ref{sec:eclipse-fitting}, simultaneously fitting for $e\cos\omega$, $S_p/S_\star$, and first order correlation with time $t$, background counts $B$, and FWHM $F$ according to Equation~\ref{eq:M_w19}. The detectability of an eclipse is derived from its marginalised eclipse depth posterior. 

Figure~\ref{fig:detectability} plots the detectability of a WASP-19b-like eclipse signal as a function of target star magnitude and eclipse depth. We find our sensitivity to eclipses plateaus for $K_\text{mag} \sim 9$ stars, for which we can detect eclipses with depths of 0.05\% at $5\sigma$ significance. The simulation is limited to stars with $K_\text{mag} > 7$. Stars brighter than $K_\text{mag}=7$ are saturated in a 1\,s exposure with IRIS2, and unlikely to yield photometry capable of secondary eclipse characterisations. WASP-19 and WASP-43 are of near-optimal brightness for the AAT+IRIS2. The deep eclipse of WASP-19b was measured at 14\,$\sigma$, whilst the WASP-43b eclipse observation, with fainter available reference stars, yielded a 7\,$\sigma$ result. Whilst our injected signals do not map the dependence of the detectability on the eclipse shape and duration, in the photon limit the detection thresholds can be scaled by the square root of the transit duration. The photometric precision is also field dependent. More precise photometry is expected for more crowded fields, until blending between neighbours becomes problematic. Using the \emph{Exoplanet Encyclopedia}\footnote{exoplanet.eu}, we identified 29 planets with $K$ band secondary eclipses potentially detectable at the $3\sigma$ level, assuming an albedo of 0.1 and a heat redistribution factor of 0.5, and observable from the Southern hemisphere. 

The high precisions delivered by the AAT+IRIS2 facility are likely due to a combination of factors. The target star was kept on the same pixel throughout the four hour eclipse observations, demonstrating the excellent tracking and guiding capabilities of the AAT. The lack of drift significantly reduces the red noise induced by inter-pixel variations and imperfect flat-fielding corrections. Since the AAT is equatorially mounted, the lack of any pupil rotation during an observation eliminates a major source of systematic error in alt-az telescopes. At the low-altitude of SSO (1165\,m), the water vapour column is consistently saturated, providing high, but stable, sky backgrounds. The large field of view, over a single detector, of IRIS2 allowed a larger selection of reference stars compared to other infrared cameras.

\section{Conclusion}
\label{sec:conclusion}

We presented new $Ks$ band secondary eclipse observations of the hot-Jupiters WASP-19b and WASP-43b, yielding eclipse depths of $0.287_{-0.020}^{+0.020}$\,\% and $0.181_{-0.027}^{+0.027}$\,\%, respectively. Our derived eclipse depths and eclipse centroids were consistent within $1\sigma$ with the majority of previous measurements. We showed that repeated observations are necessary to understand the uncertainties associated with secondary eclipse measurements, where the published errors of the ground-based observations are on average underestimated by a factor of 1.4. 

Using VSTAR atmosphere models, we examined the effects of C/O abundance, presence of clouds, and metallicity, on the theoretical spectra for these two planets. Using our observations and existing measurements, we find the atmospheres of WASP-19b and WASP-43b are marginally more consistent with carbon-rich, compared to solar, compositions, and with the addition of a top haze layer, than cloudless atmospheres. 

These observations were the first secondary eclipse measurements obtained using the AAT+IRIS2 facility. We used the observed data to demonstrate the detection capabilities of the AAT secondary eclipse programme. We find a peak photometric precision of 0.2\% at 10s cadence, allowing us to achieve $5\sigma$ detections of secondary eclipse events for eclipses deeper than 0.07\% for host stars brighter than $K_\text{mag} = 9$. We find 29 planets with detectable $K$ band eclipses potentially observable using the AAT.

\section*{Acknowledgements}
\label{sec:acknowledgements}
This research was supported by ARC Super Science Fellowship grant FS100100046 and ARC Discovery grant DP130102695.

\bibliographystyle{mn2e}
\bibliography{mybibfile}

%\FloatBarrier

\label{lastpage}

\end{document}